\documentclass[twoside]{dis09}
\usepackage[latin1]{inputenc}
\usepackage[dvips]{graphicx,epsfig,color}
\usepackage{wrapfig,rotating}
\usepackage{amssymb,amsmath,array}

\begin{document}
\title{Top Production Cross Sections at D{\O}}

\author{Ji\v{r}\'{\i} Kvita
\thanks{The author would like to acknowledge the support of the research plan MSM0021620859 of the Ministry of Education, Youth and Sports of the Czech Republic}
\vspace{.3cm}\\
Institute of Particle and Nuclear Physics, \\ Faculty of Mathematics and Physics, Charles University in Prague \\
V Hole\v{s}ovi\v{c}k\'{a}ch 2, 180~00 Prague, Czech Republic
}

\maketitle


\begin{abstract}
 We report on measurements of the $t\bar{t}$ production cross section at a center-of-mass energy of 1.96 TeV at the~D{\O} experiment during Run II of the Fermilab Tevatron collider. We use candidate events in lepton+jets and dilepton final states. In the most sensitive channel (lepton+jets channel), a neural network algorithm that uses lifetime information to identify b-quark jets is used to distinguish signal from background processes. We also present measurements of single top quark production at D{\O} using several multivariate techniques to separate signal from background.
\end{abstract}


\section{Introduction}

Top quark is the heaviest elementary particle known to-date discovered at Fermilab in 1995 \cite{cdf_ttbar_observation,dzero_ttbar_observation}. During the current Run~II period of the Tevatron accelerator, the growing statistics permits to study in detail further kinematic properties of the top pair production as well as recently observed electroweak production of individual top quarks. We preset the latest results on the $t\bar{t}$ cross sections measurements, their combination and interpretation in terms of limits on new physics. We summarise the current state of the observation of the single top quark production and the measurements and searches based on the $Wtb$ vertex properties.


\section{$t\bar{t}$ Production Cross-section}

As the top quark was first observed produced in $t\bar{t}$ pairs via the strong interaction, the pair production is the obvious first handle to test perturbative quantum chromodynamics (pQCD) while the decaying system properties can be used as an electroweak laboratory to search for new physics. At Tevatron the production proceeds mostly via $q\bar{q}$ annihilation (85\%) accompanied by the $gg$ fusion (15\%).
The final states of the $t\bar{t}$ system are determined by the leptonic, hadronic or mixed decays modes of the $W$-bosons.
The dilepton ($\ell\ell$) final state is easy to identify, has low background, but also low branching fraction; the $\ell+$jet channels (with prompt $W\rightarrow e,\,\mu$ or cascade $\tau \rightarrow e,\,\mu$) have larger branching fraction, more background, but have only one escaping neutrino; the hadronic channel has the largest branching ratio, but suffers from overwhelming background from QCD multijet production. Dedicated $\tau+$jets analyses are challenged by the $\tau$ lepton identification, but provide important additional information and offer windows on testing new models with prominent couplings to $\tau$.

\begin{figure}[h]
\centerline{
\includegraphics[width=0.4\textwidth,angle=0,clip=,]{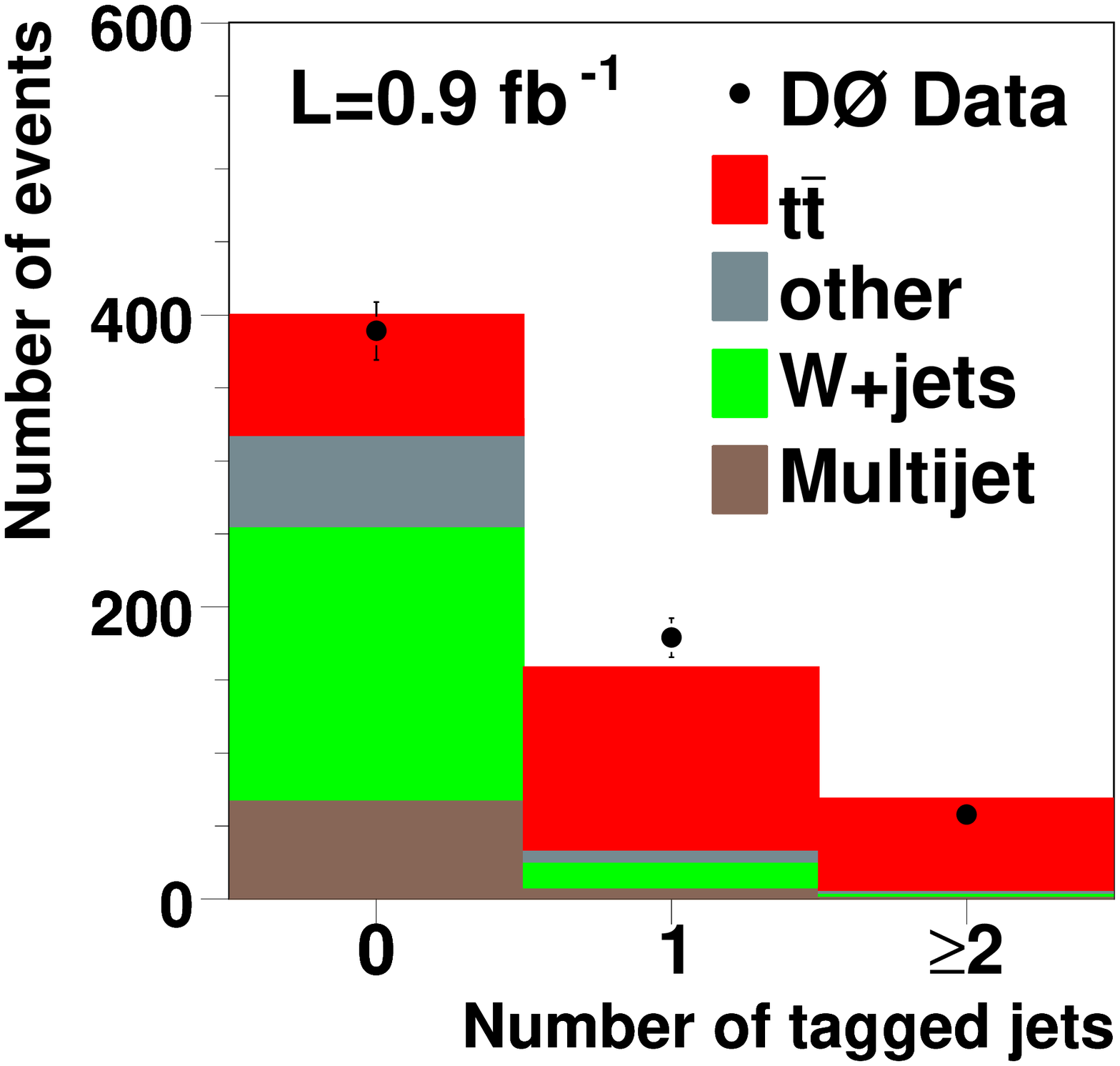}
\includegraphics[width=0.4\textwidth,angle=0,clip=,]{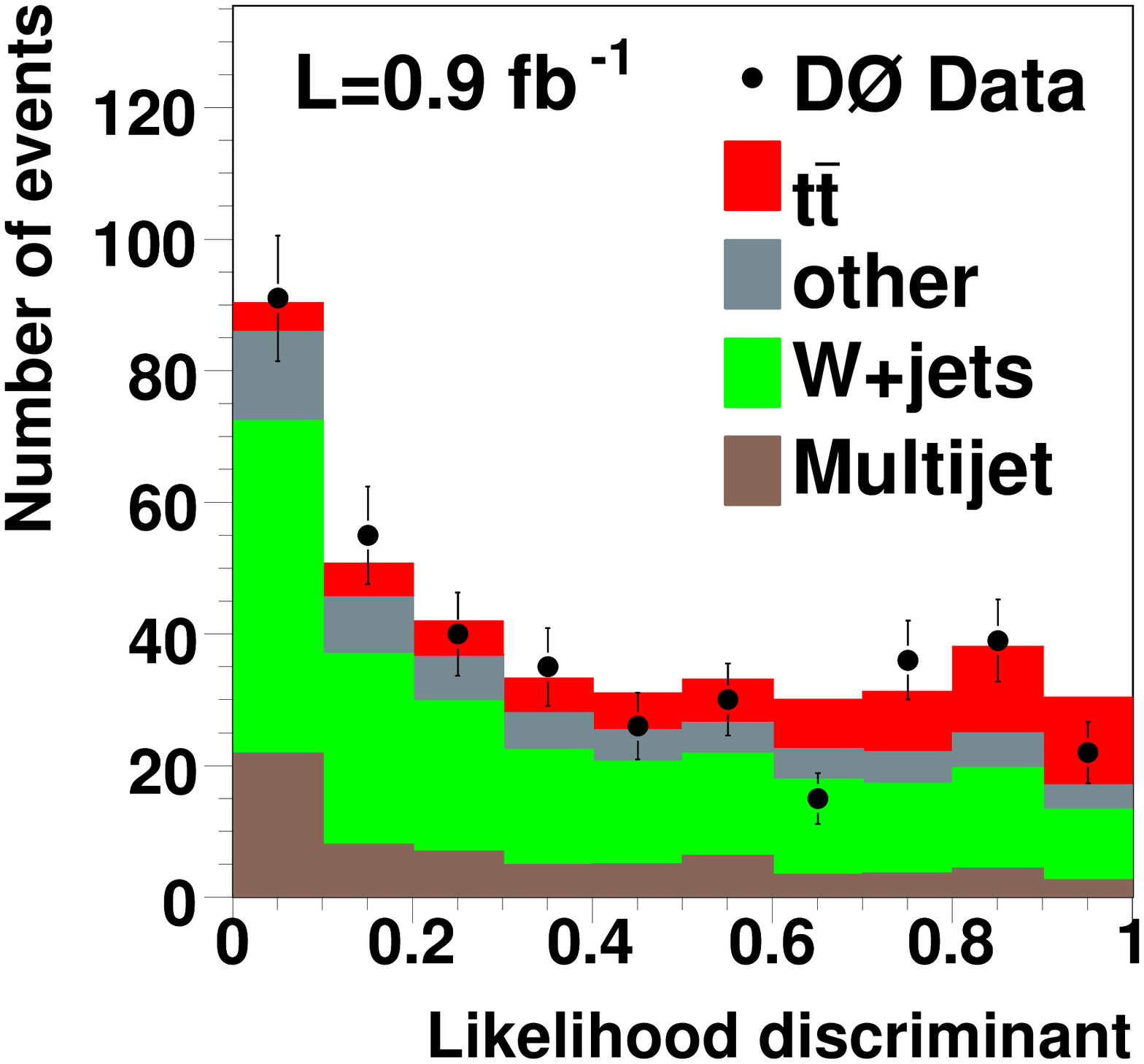}
}
\caption{Number of $b$-tagged jets in the 4-jet bin in $t\bar{t}\rightarrow \ell+$jets (left) and the likelihood discriminant used to fit the signal fraction (right).}\label{Fig:ljjets}
\end{figure}

The cross-section analysis in the $\ell+$jets was divided into several bins in terms of the jet multiplicity, with $\ell$+1,2~jets serving as a control sample to tune selection and data/simulation agreement, and the 3 and $\geq 4$ jet bins (with larger signal fraction) for the measurement itself. We required at least one jet identified as a $b$-jet using a Neural Network (NN) $b$-tagger, one isolated lepton and large missing transverse energy.  Additionally  the leading jet $p_T$ was required to be $\geq 40\,$GeV. Most of Standard Model backgrounds were modelled using {\sc Alpgen}+{\sc Pythia} parton shower, including theory NLO/LO scale factors. The largest physics background comes from the $W/Z+$jets events, other backgrounds include diboson and single top production. As there is no reliable model of the multijet production beyond several final state hard partons, the multijet background was taken from a special data with inverted cut on the lepton isolation. 
The signal fraction grows with the number of the $b$-tagged jets (see~Figure~\ref{Fig:ljjets}, left). A kinematic likelihood discriminant based on several topological and kinematic variables (see~Figure~\ref{Fig:ljjets}, right) was employed to fit the $t\bar{t}$ fraction and the cross-section.

Cross-section measurements in $\ell+$jets, $\ell\ell$ and $\tau\ell$ channels were combined using a total of {14 orthogonal channels}, the combined cross-section was fitted using a likelihood function with Poisson terms for signal and background yields. Systematic uncertainties are included using Gaussian probability density function. The combined result is
$ \sigma_{t\bar{t}} = 8.18^{+0.98}_{-0.87}\,{\rm pb}$, assuming $m_{\rm top} = 170\,{\rm GeV}$ \cite{dzero_cross_section_combination}. The systematic uncertainties dominate the error on this result.

\begin{figure}[h]
\centerline{
\includegraphics[width=0.450\textwidth,angle=0,clip=,]{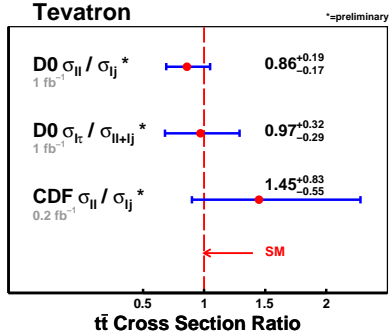}
}
\caption{Ratios of cross-section as measured by the D{\O} and CDF compared to the SM.}
\label{Fig:xsect_combi_d0_cdf}
\end{figure}

We also provide the ratios of cross-sections between the channels $R_\sigma$, where most of systematics cancel, leading to results on $R_{\ell\ell/\ell{\rm j}} = 0.86^{+0.19}_{-0.17}$ and  $R_{\tau\ell/\ell\ell+\ell{\rm j}} = 0.97^{+0.32}_{-0.29}\,,$
which are consistent with the SM value of 1 and are shown in~Figure~\ref{Fig:xsect_combi_d0_cdf}.

\begin{figure}[t]
\centerline{
\includegraphics[width=0.45\textwidth,angle=0,clip=,]{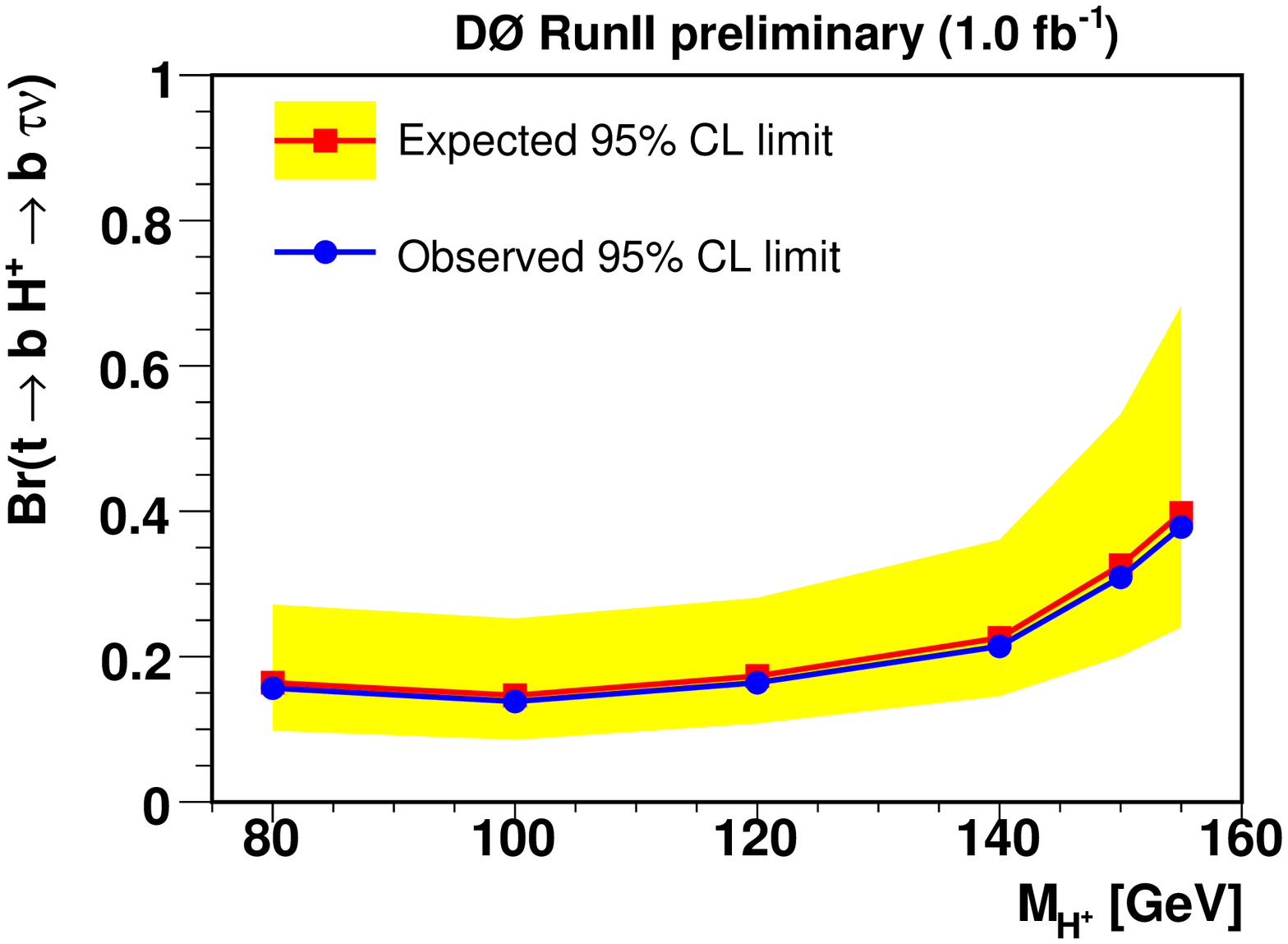}
\includegraphics[width=0.45\textwidth,angle=0,clip=,]{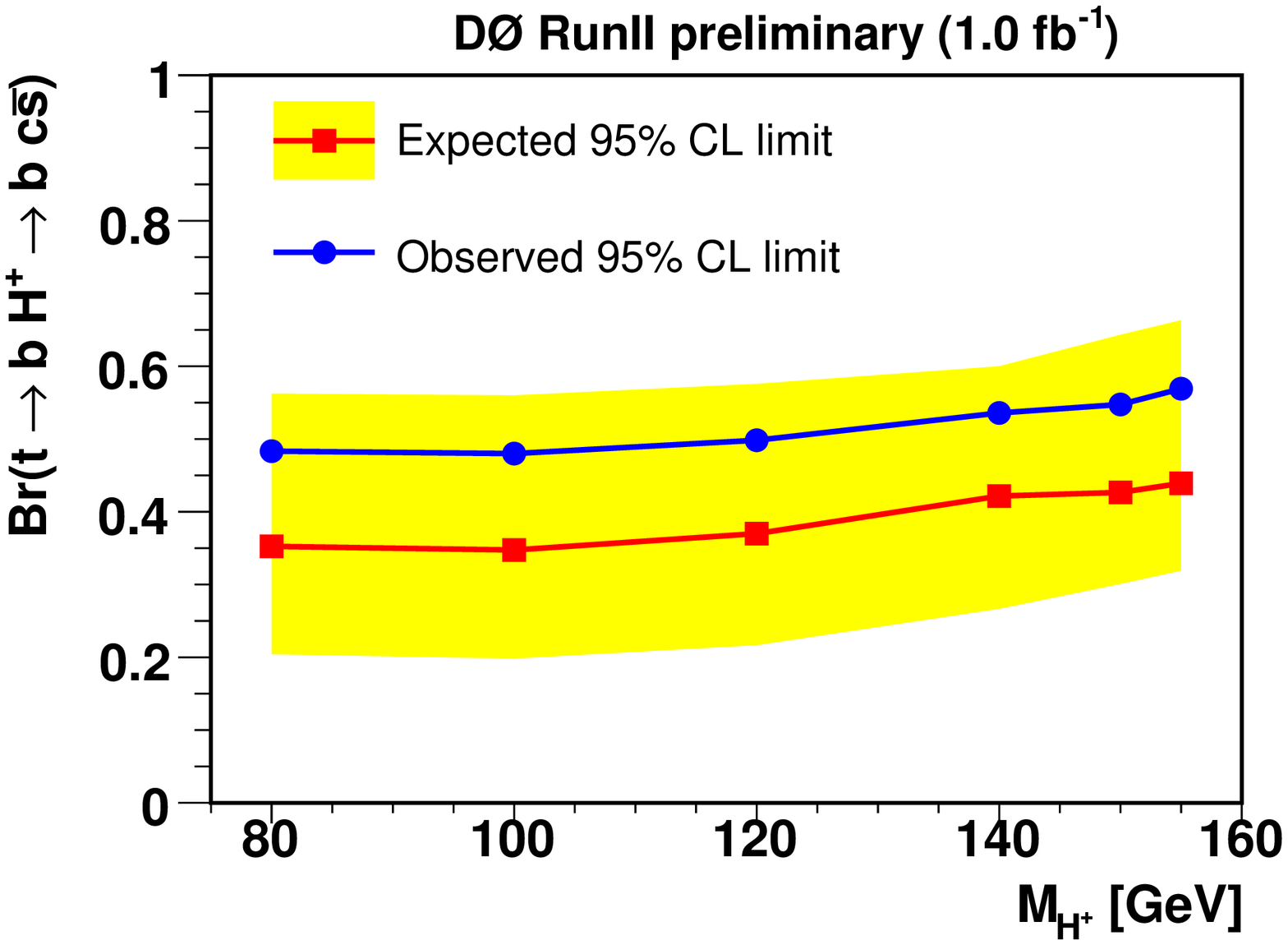}
}
\caption{Upper limits on $B(t \rightarrow H^{+}b)$ for tauonic (left) and leptophobic (right) $H^{+}$ decays. The yellow band shows the 1-sigma band around the expected limit.}
\label{Fig:ttbar_H_limits}
\end{figure}

In addition, cross-section ratios between individual channels can be used to set upper limits on the $B(t\rightarrow H^+ b)$ branching ratio, as for $H^+ \rightarrow \tau\nu_\tau$ one would observe an enhancement of the $\tau$ channel (using $R_{\tau\ell/\ell\ell+\ell{\rm j}}$). Leptophobic $H^+ \rightarrow c\bar{s}\,\,\,\,$  would lead to an enhancement of the $\ell+$jets channel (using $R_{\ell\ell/\ell{\rm j}})$. We extracted the expected and observed upper limits as shown in~Figure~\ref{Fig:ttbar_H_limits} using pseudo-experiments derived from the data distributions.

\begin{figure}[!h]
\centerline{
\includegraphics[width=0.65\textwidth,angle=0,clip=,]{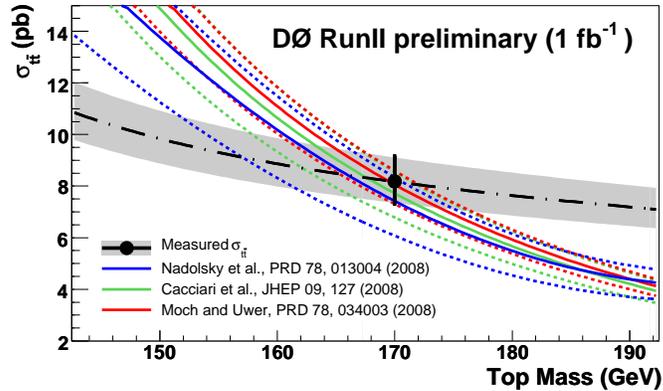}
}
\caption{The dependence of the theoretical and measured $t\bar{t}$ cross sections on the top quark mass, used for the top mass extraction from the measured cross-section.}
\label{Fig:mass_from_xsect}
\end{figure}

The current Tevatron $m_{\rm top}$ combined measurement is precise to $1.3\,$GeV \cite{mtop_world_aver}, triggering discussions on the definition and interpretations of the top mass which is actually measured. A complementary information on $m_{\rm top}$ with different sensitivity to theoretical and experimental uncertainties can be extracted from the $\sigma_{t\bar{t}}(m_t)$ dependence.  Assuming different $m_t$, the measured $t\bar{t}$ cross-section can be compared to NLO and approximate NNLO (NLL) calculations, 
where $m_{\rm top}$ definition is usually the pole mass. The extracted values of $m_{\rm top}$ range from $165.5$ to $169.1\,$GeV depending on the theory used, with errors of $\approx 6\,$GeV, being in good agreement with the world-average of $173.1\pm 1.3\,$GeV \cite{mtop_world_aver} as depicted in~Figure~\ref{Fig:mass_from_xsect}.

\section{Single Top Quark Observation}

The electroweak (single) top quark production had been one of the remaining SM processes to be observed until recently \cite{dzero_single_top_observation}, with the evidences for single top production reported by the Tevatron collider experiments in 2007 \cite{dzero_single_top_evidence} and 2008 \cite{cdf_single_top_evidence}. Indeed, this process has a much larger background and smaller cross-section compared to the $t\bar{t}$ production. At the same time, observation of the single top production is rewarding in providing a direct access to the $Wtb$ vertex properties and another window to new physics. Two basic production modes can be distinguished at LO, denoted the $s$ and $t$-channels.


  The basic experimental requirements to measure the single top cross-section include the selection of events with one isolated lepton and at least two jets with at least one with $p_T > 25\,$GeV. In addition, the missing transverse energy ($\nu$-signature) is required to be $> 20(25)\,$GeV in the two-jet ($\geq 3$) bin. One or two of the jets is required to originate from $b$-hadrons using the secondary vertex information incorporated into a Neural Net $b$-tagger. Selecting 4,519 $b$-tagged events, we expect $223\pm 30$ single top events with signal acceptances of 3.7\% (2.5\%) for $tb$ ($tbq$). As the expected signal fraction of 3--9\% is in fact smaller than the background uncertainty, a simple counting experiment method is impossible and multivariate techniques are needed to extract the signal.

\begin{figure}[h]
\centerline{
 \includegraphics[width=0.4\textwidth,angle=0,clip=,]{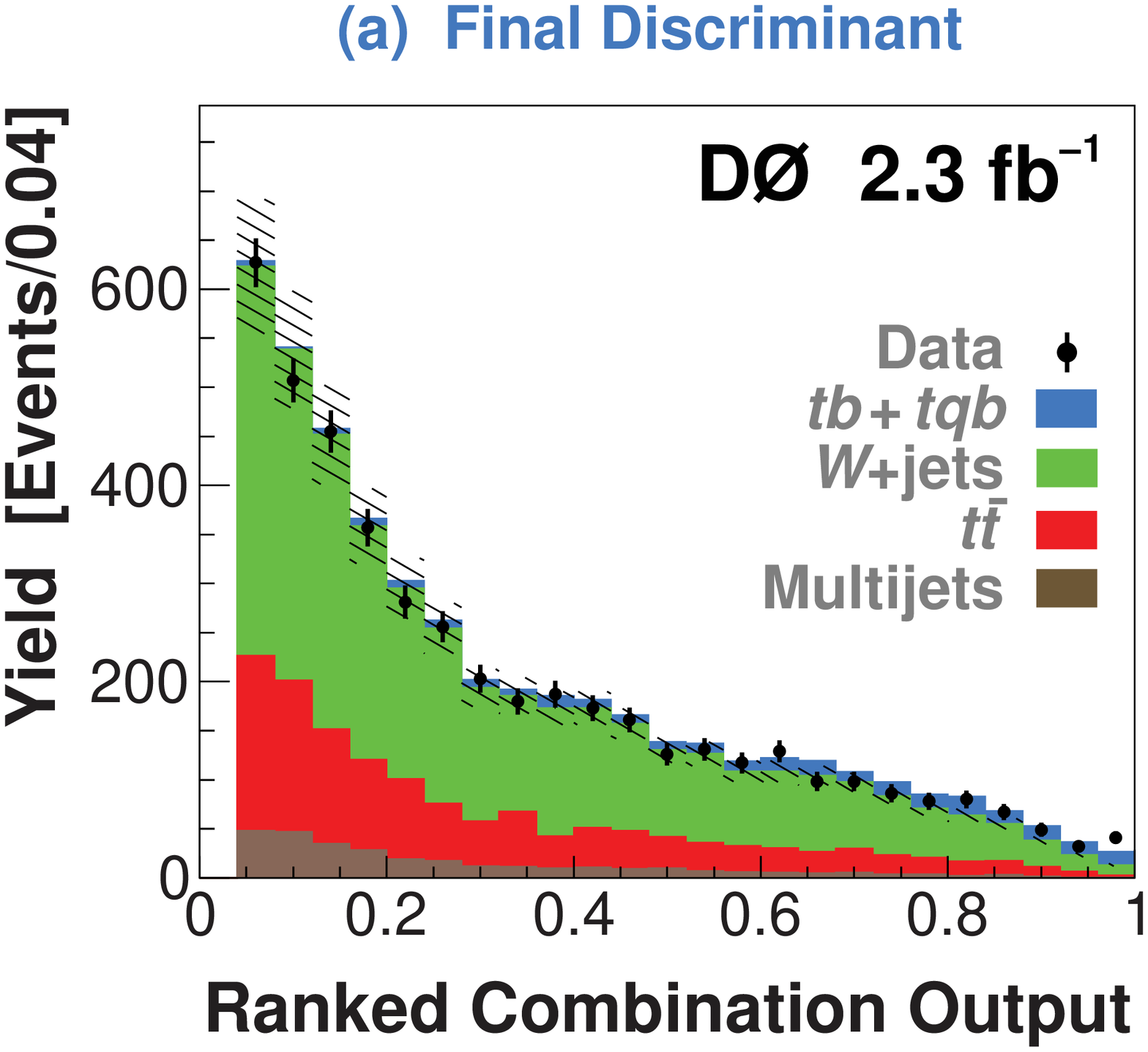}
  \includegraphics[width=0.4\textwidth,angle=0,clip=,]{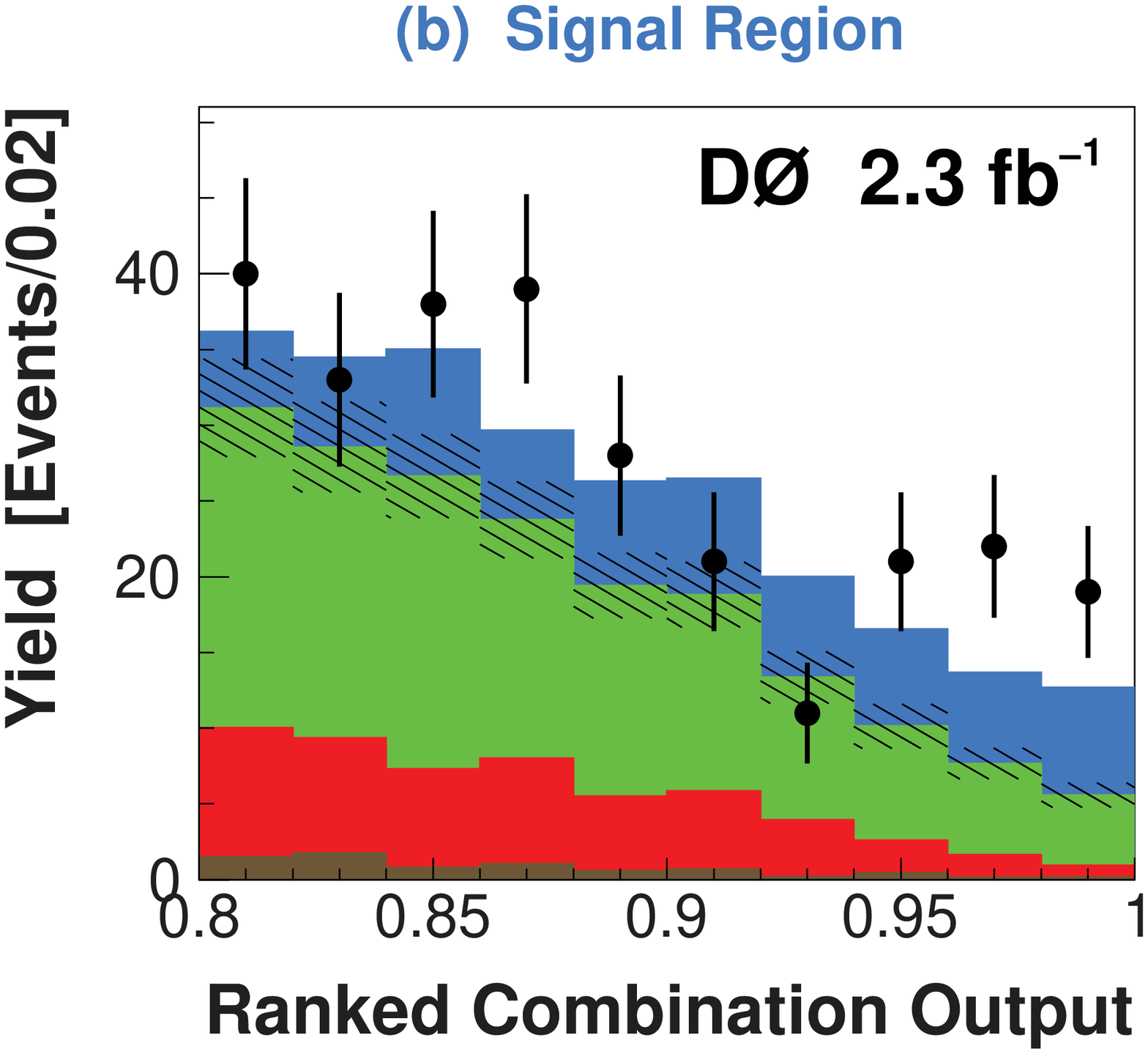}
}
\caption{Combined output of the three discriminants used to distinguish between single top (peaking on the right) and  and background (peaking on the left).}
\label{Fig:d0_singletop_2}
\end{figure}

Three analysis methods named Boosted Decision Trees (BDT), Bayesian Neural-Network (BNN), Matrix Element (ME) giving consistent results were employed and used to produce a combined discriminant (see~Figure~\ref{Fig:d0_singletop_2}). The result based on $2.3\,{\rm fb}^{-1}$ data collected by the D{\O} detector is shown in~Table~\ref{Table:singletop_individual}. Being 50--60\% correlated, combining the individual results leads to an improved significance. The combined cross-section measurement is then $\sigma (p\bar{p} \rightarrow tb + X,\, tbq + X) = 3.94\pm 0.88\,{\rm pb}\,.$ 
The probability to measure such cross-section or higher in absence of signal is $2.5\times 10^{-5}$, corresponding to $5.0\,\sigma$ significance. 
\begin{table}[h]
\begin{center}
  \begin{tabular}{|l|c|cc|} \hline 
           &               & \multicolumn{2}{c|}{Significance\hfill} \\ 
    Method & Cross-section & Expected & Observed \\ \hline
    BDT & $ {3.74}^{+0.95}_{-0.79} \,{\rm pb}$ & 4.3 & 4.6\\
    BNN & $ {4.70}^{+1.18}_{-0.93} \,{\rm pb}$ & 4.1 & 5.4 \\
    ME  & $ {4.30}^{+0.99}_{-1.20} \,{\rm pb}$ & 4.1 & 4.9 \\  \hline 
  \end{tabular}
\end{center}
\caption{Single top quark cross sections measured using three different methods.}
\label{Table:singletop_individual}
\end{table}


\begin{figure}[h]
\centerline{
\includegraphics[width=0.32\textwidth,angle=0,clip=,]{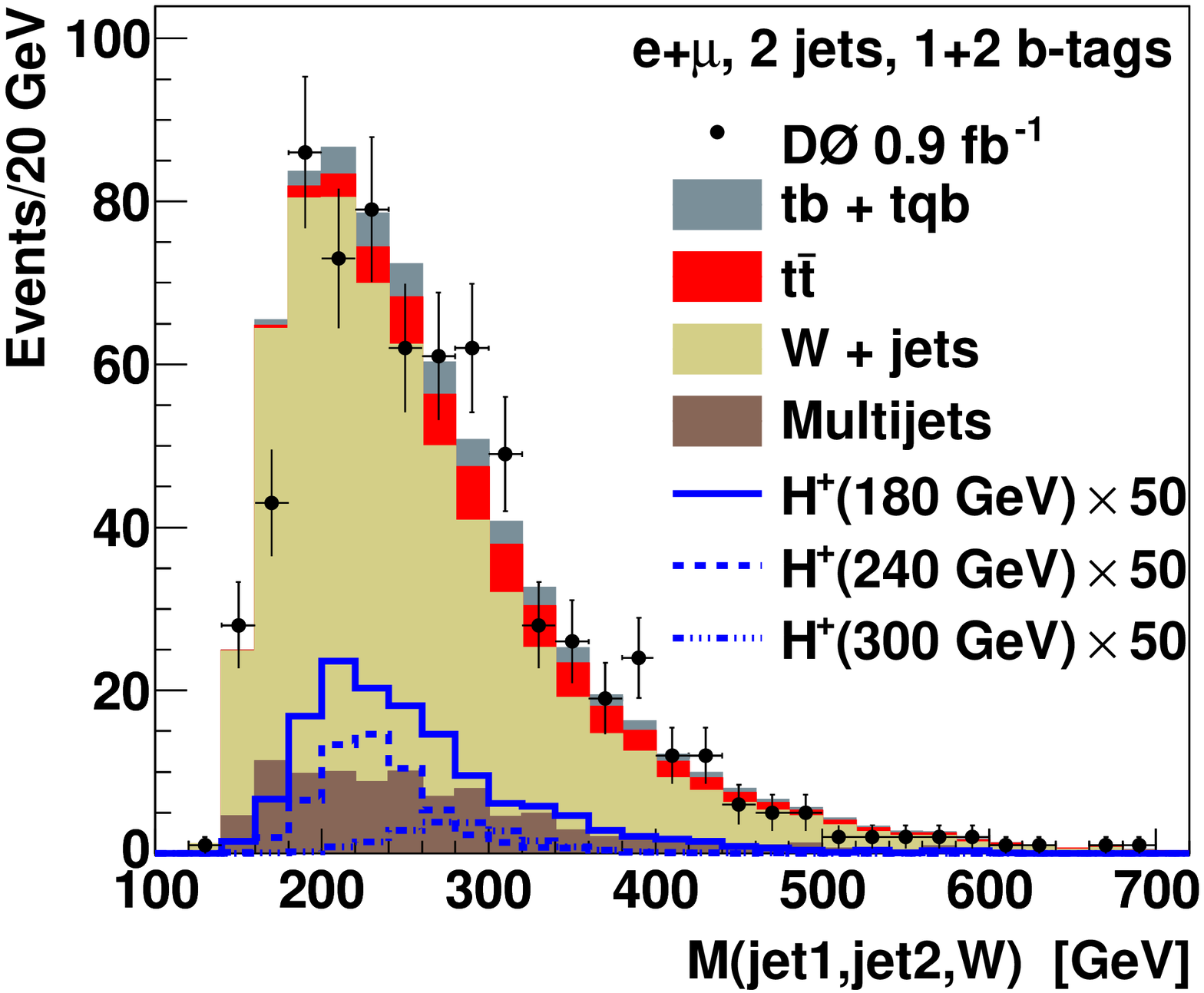}
\includegraphics[width=0.40\textwidth,angle=0,clip=,]{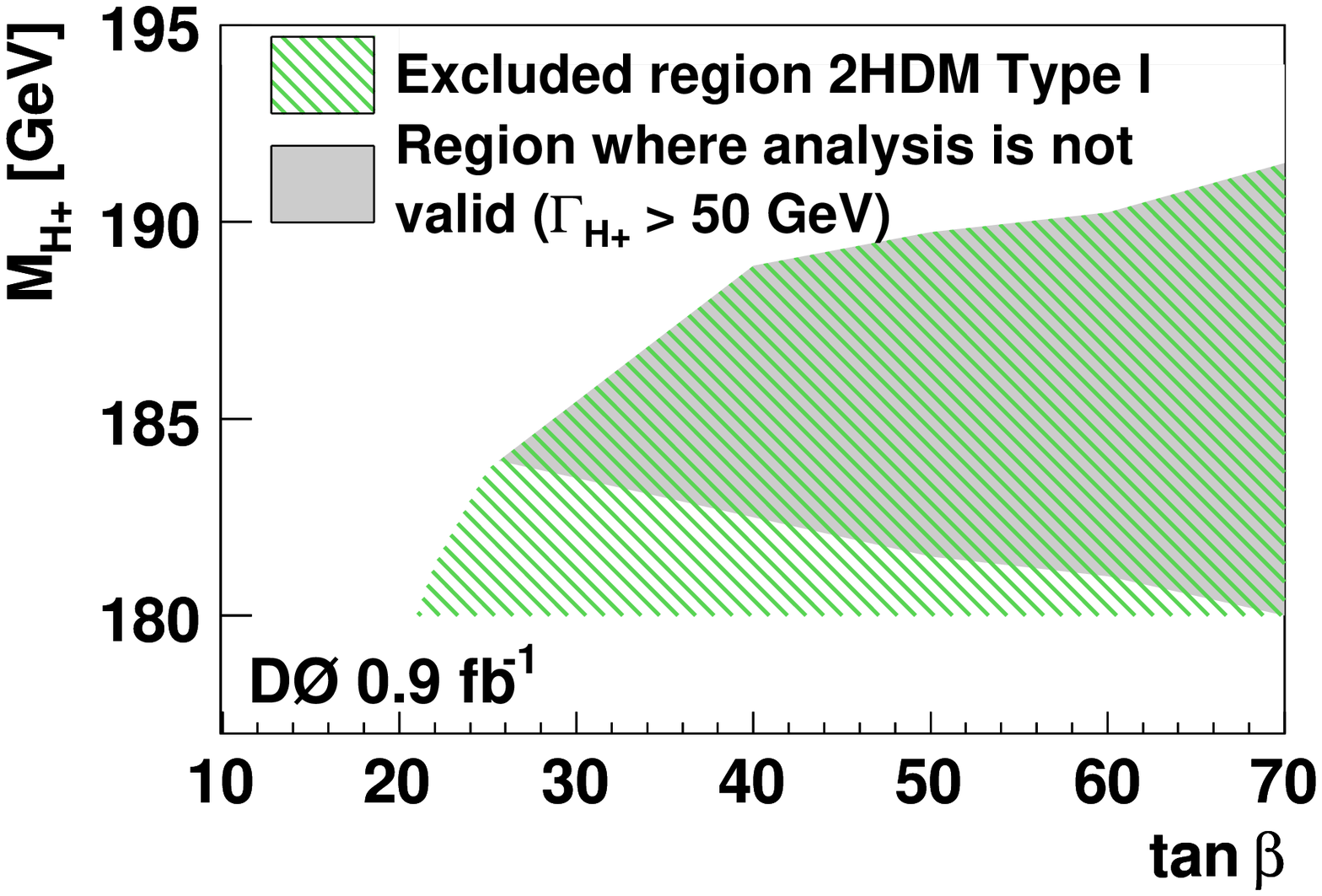}
}
\caption{Jet-jet-$W$ candidates invariant mass (left) and limits on the charged Higgs $H^+$ mass (right) in the $tb$ final state}
\label{Fig:d0_singletop_4}
\end{figure}
One can also search new phenomena in the same final state, e.g. for charged Higgs bosons decaying to top and bottom quarks in $p\bar{p}$ collisions \cite{d0_paper_single_top_higgs}. Limits on the charged Higgs mass a function of $\tan\beta$ were set based on a discriminating variable of the jet-jet-$W$ invariant mass (Figure~\ref{Fig:d0_singletop_4}). 
Single top production also provides a direct access to the properties of the $Wtb$ vertex, allowing a search for anomalous top quark couplings \cite{d0_paper_single_top_VA}. The SM Lagrangian allows only the $V-A$ vertex structure, but a more general form can be probed:
  $$ \mathcal{L} = -\frac{g}{\sqrt{2}} \bar{b}\gamma^\mu V_{\rm tb} \left(  f_1^L \mathcal{P}_L + f_1^R \mathcal{P}_R\right)\,t W_\mu^-    -\frac{g}{\sqrt{2}} \bar{b}\,{\rm i}\sigma^{\mu\nu}q_\nu  V_{\rm tb} \left(  f_2^L \mathcal{P}_L + f_2^R \mathcal{P}_R\right)\,t W_\mu^- + {\rm h.c.}\,,$$
where SM corresponds to the case of $f_1^L = 1$, $f_1^R = f_2^L = f_2^L = 0$. 
Using the $W$ helicity measurement in the $t\bar{t} \rightarrow \ell+$jet channel (which constrains form factors ratios) for probability density priors, the extracted Bayesian limits on the vector and axial vector form factors are $|f_1^R|^2 < 1.01$, $|f_2^L|^2 < 0.28$ and $|f_2^R|^2 < 0.23$ assuming $f_1^L = 1$.



\begin{footnotesize}


\end{footnotesize}



\begin{thebibliography}{99}
\bibitem{url_ID_153_53294_Session_5}
\verb$http://indico.cern.ch/materialDisplay.py?contribId=153&sessionId=5&materialId=slides&confId=53294$.

\bibitem{cdf_ttbar_observation} Phys.~Rev.~Lett., Vol. {\bf 74}, No. 14. (April 1995), 2626.
\bibitem{dzero_ttbar_observation} Phys.~Rev.~Lett., Vol. {\bf 74}, No. 14. (Apr 1995), 2632.

\bibitem{dzero_cross_section_combination} {\tt arXiv:0903.5525v1 [hep-ex]}, {\tt FERMILAB-PUB-09-092-E}; submitted to Phys.~Rev.~D.
\bibitem{mtop_world_aver} {\tt arXiv:0903.2503 [hep-ex]}, {\tt FERMILAB-TM-2427-E}.

\bibitem{dzero_single_top_observation} {\tt arXiv:0903.0850v1 [hep-ex]}, {\tt FERMILAB-PUB-09-056-E}.
\bibitem{dzero_single_top_evidence} Phys.~Rev.~Lett. {\bf 98}:181802 (2007), {\tt arXiv:0612052v2 [hep-ex]}.

\bibitem{cdf_single_top_evidence} Phys.~Rev.~Lett. {\bf 101}:252001 (2008), {\tt arXiv:0809.2581 [hep-ex]}.

\bibitem{d0_paper_single_top_higgs} Phys.~Rev.~Lett. {\bf 102}:191802 (2009), {\tt arXiv:0807.0859 [hep-ex]}.

\bibitem{d0_paper_single_top_VA} Phys.~Rev.~Lett. {\bf 102}:092002 (2009), {\tt arXiv:0901.0151 [hep-ex]}.

\end{thebibliography}
\end{document}